\documentclass[useAMS,usenatbib]{mn2e}

\usepackage{graphicx}
\usepackage{dcolumn}
\usepackage{bm}
\usepackage{amsmath}
\usepackage{amssymb}

\title[Calculation of rate constants for vibrational and rotational excitation of the $\mathrm{H}_3^+$ ion by electron impact]{Calculation of rate constants for vibrational and rotational excitation of the $\mathrm{H}_3^+$ ion by electron impact}
\author[V.~Kokoouline, A.~Faure, J. Tennyson, and C. H. Greene]{Viatcheslav Kokoouline$^1$\thanks{E-mail:
slavako@mail.ucf.edu}, Alexandre Faure$^2$, Jonathan Tennyson$^3$, Chris H. Greene$^4$\\
$^1$Department of Physics, University of Central Florida, Orlando, Florida 32816, USA\\
Laboratoire Aim\'e Cotton, CNRS, Universit\'e Paris 11, 91405 Orsay France;\\
$^2$ Laboratoire d'Astrophysique, UMR 5571 CNRS, Universit\'e Joseph-Fourier, B.P. 53, 38041 Grenoble cedex 09, France;\\
$^3$ Department of Physics and Astronomy, University College London, Gower Street, London WC1E 6BT, UK;\\
$^4$ Department of Physics and JILA, University of Colorado, Boulder, Colorado 80309, USA}
\begin{document}


\pagerange{\pageref{firstpage}--\pageref{lastpage}} \pubyear{2010}

\maketitle

\label{firstpage}

\begin{abstract}
We present theoretical thermally-averaged rate constants for
vibrational and rotational (de$-$)excitation of the H$_3^+$ ion by
electron impact. The constants are calculated using the multi-channel
quantum-defect approach. The calculation includes processes
that involve a change $|\Delta J|\leq 2$ in the rotational angular
momentum $J$ of H$_3^+$.  The rate constants are calculated for states
with $J\leq 5$ for rotational transitions of the H$_3^+$ ground
vibrational level. The thermal rates for transitions
among the lowest eight vibrational levels are also presented, averaged over the
rotational structure of the vibrational levels. The conditions for
producing non-thermal rotational and vibrational distributions of
H$_3^+$ in astrophysical environments are discussed.
\end{abstract}


\section[]{Introduction}

Rotational and vibrational excitation of small polyatomic ions by
electron impact is one of the important processes occurring in a
neutral molecular plasma. In particular, the probability of
rovibrational (de-)excitation in electron-ion collisions can be
relatively high. The high probabilities and correspondingly high rate
constants are driven by the non-Born-Oppenheimer coupling between
electronic and rovibrational motions of the ion-electron system. In
certain small polyatomic molecules, the coupling is particularly
strong, as is the case for the H$_3^+$ ion.

Owing to its importance in interstellar space \citep{oka06b},
planetary ionospheres \citep{miller00} and laboratory experiments
\citep{larsson00,plasil02,johnsen05}, the H$_3^+$ ion has been studied
for many years.  In particular, processes involving electron
scattering from the ion have been recently studied experimentally and
theoretically. Such processes include electron-impact rovibrational
excitation \citep{faure02,faure06a,faure09}, dissociative
recombination \citep{larsson00,kokoouline01b,johnsen05,santos07},
electronic excitation and ionisation \citep{gorfinkiel04} and
photoionization
\citep{bordas91,stephens94,stephens95,mistrik00,kokoouline04a} of the
metastable neutral H$_3$ molecule.  Astrophysically, electron impact
excitation of molecular ions has been observed to be the dominant
collisional excitation process in some environments
(e.g. \citealt{serra06}).

Previously, cross sections and rate constants for a few rotational
transitions in H$_3^+$ have been calculated
\citep{faure02,faure03,kokoouline03b,faure06a,faure06b}. Vibrational
rate constants have apparently not been studied extensively for this
fundamental ion.  Here, we present thermal rate constants for
transitions between different rotational states of the ground
vibrational level of H$_3^+$ with low angular momentum, $J\le 5$. We
also present thermal rate constants for rotationally-averaged
transitions between different vibrational levels.

The next section of the article briefly discusses the theoretical
approach used in the present calculation. A detailed description of
the approach is lengthy and has already been published elsewhere
\citep{kokoouline03b,santos07}. Therefore we only sketch here the main ideas
of the approach. In section \ref{sec:vib_exc}, we present the rates
for vibrational (de-)excitation of H$_3^+$. Section \ref{sec:rot_exc}
is devoted to the calculation of rotational rate constants for
transitions within rotational manifold of the ground vibrational
level. Astrophysical implications are discussed in
Section~\ref{sec:astro}. Finally, Section \ref{sec:concl} presents our
conclusions.

\section{Theoretical approach}
\label{sec:theory}

The theoretical model employed in the present study is based on quantum
defect theory. It is discussed in detail by
\cite{kokoouline03a}, \cite{kokoouline03b}, \cite{faure06a} and
\cite{santos07}. Here, we only
mention the main ideas used in the model.

The energy-dependent theoretical rate constant $\alpha_{i'\leftarrow
i}(E)$ for a transition from the initial rovibrational level $i$ to a
final one $i'$ is obtained from the corresponding matrix element
$S^{phys}_{i'i}(E)$ of the energy-dependent scattering operator, $\hat
S^{phys}(E)$.  The main difficulty in the theoretical approach is the
construction of the scattering matrix $S^{phys}_{i'i}(E)$, (the indices $i$ and $i'$
 refer to specific rovibrational states of the H$_3^+$ 
ion; the electron angular momentum and its coupling with the ion to form 
a total angular momentum eigenstate are implied as well, but these will 
be suppressed in our notation since they are diagonal quantum numbers in 
the present approximation).  For
our discussion, it is convenient to represent the index $i$ as $rv$, where
$r=(JK)$ specifies the rotational quantum numbers, i.e. ionic angular
momentum $J$ and its projection $K$ on the molecular axis; and
$v=\{v_1,v_2^{l_2}\}$ specifies the vibrational quantum numbers in a
normal mode classification. We note that our model neglects the explicit
coupling between rotational and vibrational angular momenta which occurs
for excited vibrational states with $l_2 > 0$ and which leads to a more
complicated set of quantum numbers \citep{lindsay01}.

The construction of the scattering matrix $S^{phys}_{i'i}(E)$ begins
from the {\it ab initio} potential surfaces of the ground electronic
state of the ion and several excited states of the neutral H$_3$
molecule \citep{mistrik00}, ($U^+({\cal Q})$ for the ion and
$U_n({\cal Q})$ for the neutral molecule).  We will use the symbol
$\cal Q$ to specify collectively the three internuclear distances. For
a given geometry $\cal Q$, the electronic wave function of the outer
electron of the H$_3$ excited states resembles the electronic wave
function of the hydrogen atom. However, due to fact that at short
distances from the ionic core the electron-ion interaction is
different from that in the hydrogen atom, the electron binding energy
$U^+({\cal Q})-U_n({\cal Q})$ is generally different than the
corresponding binding energy $1/(2n^2)$ in the hydrogen atom ($n$ is
the principal quantum number). The departure of the $U^+({\cal
  Q})-U_n({\cal Q})$ difference from $1/(2n^2)$ is written as
$1/(2(n-\mu)^2)$, where the quantum defect $\mu$ is only weakly
dependent on the principal quantum number $n$. When the energy $E$ of
the electron+ion system approaches the ionization limit ($n\to
\infty$) and becomes larger, the principal quantum number $n$ looses
its physical meaning, but the quantum defect $\mu$ does not: it gives
the collisional phase shift $\delta(E)=\pi\mu$ in terms of the
scattering phase in electron-ion collisions at energies above the
ionization limit \citep{seaton66}. The phase shift also depends weakly
on the energy $E$ and it determines the scattering matrix
$S(E)=\exp{(2i\delta)}$. This is the reason why the energies of
excited electronic states in the neutral molecule can be used to
obtain the quantum defect and to describe collisions between the ion
and the electron.

The preceding discussion assumes that the electron scatters from (or is
bound to) the molecular ion, which stays at a given geometry, i.e. the
nuclei remain fixed throughout. This approximation is only reliable on a
time scale much shorter than the period of ionic vibrational motion. In
this limit, the quantum defects are functions of geometry $\cal Q$,
$\mu({\cal Q})$. Generally speaking, the dependence of $\mu({\cal Q})$ is
much smoother than the dependence of $U^+({\cal Q})$ and $U({\cal Q})$.
Although the quantum defect depends only weakly on the principal quantum
number $n$, it usually depends strongly on the angular momentum $l$ and on
its projection $\Lambda$ on the molecular axis of the ionic core. For a
nonlinear triatomic ion such as H$_3^+$, the body-frame quantization axis
is chosen as the normal to the plane containing the nuclei. At large $l\ge
2$, the quantum defects become small, because the electronic wave function
of high $l$ approaches the unperturbed wave function in the hydrogen atom.

Therefore, for the description of an electron-ion scattering process
involving low values of $l$, one must obtain the quantum defect
functions $\mu_{l\Lambda}({\cal Q})$ for various $l$ and $\Lambda$ and
the corresponding scattering matrices $S({\cal
  Q})=\exp{(2i\delta_{l\Lambda})}$. The calculated potential energy
surfaces are obtained in the Born-Oppenheimer approximation
\citep{mistrik00}, i.e. in which the coupling between electronic and
nuclear motion is neglected. The quantum defects and the scattering
matrices obtained from the potential surfaces as described above,
therefore, fail to account for the coupling. However, in the H$_3^+
+e^-$ case a strong non-adiabatic Jahn-Teller interaction must be
accounted for in order to appropriately describe the scattering
process. It mixes the two $\pi$ electronic states of the same orbital
angular momentum $l$ of the incoming electron and can be included in
the scattering matrix built from quantum defects using the formalism
suggested in \cite{staib90}. The resulting scattering matrix is now
not diagonal over the $\Lambda$ quantum numbers. The strongest
coupling is between the $p\pi$ electronic states. In our treatment we
include only $p$-states ($\pi$ and $\sigma$). This implies that this
body-frame scattering matrix $S^{BF}_{\Lambda',\Lambda}({\cal Q})$ is
a $3\times 3$ matrix \citep{kokoouline03b}.

The scattering matrix constructed in this way represents the electron-ion
scattering only if the ion stays at the same configuration ${\cal Q}$
during the entire process, which is not a valid description.  The
physically meaningful scattering matrix must describe the amplitude of
scattering from a particular rovibrational state of the ion to another
one, including the possibility of nuclear motion during the full collision
process. We denote this as the space-fixed scattering matrix
$S_{s',s}^{SF}$,  where $s$ and $s'$ refer to initial and final states of
the ion. The formalism of the rovibrational frame transformation 
\citep{atabek74,jungen77,fano75} allows us to use the matrix $S^{BF}({\cal
Q})$ to construct the $S_{s',s}^{SF}$ matrix. In this formalism, the two
matrices are considered as two equivalent forms of the same scattering
operator in two different representation bases. The representation basis
of $S_{s',s}^{SF}=\langle s'|\hat S|s\rangle$ is the set of rovibrational
energy eigenstates  $|s\rangle$. The basis for the $S^{BF}({\cal Q})$
matrix is made of tensor products $|b\rangle=|{\cal
Q}\rangle|\Lambda\rangle$, where $|{\cal Q}\rangle$ represents the
vibrational position eigenstates of the ion, $|\Lambda\rangle$ is the
angular state vector of the electron in either the $p\pi$ or $p\sigma$
state.

The two representations are connected by the standard basis transformation
formula
\begin{eqnarray}
\label{eq:RVFT}
S^{SF}_{s';s}=\sum_{b',b} \langle s'|b'\rangle  \langle b'|\hat S|b\rangle
\langle b|s\rangle\,,
\end{eqnarray}
where the summation indicates a sum over discrete indices and an
integration over the continuous coordinates $\cal Q$ and rotational
coordinates (three Euler angles). Because the scattering matrix in the
$|b\rangle$ basis is diagonal over $|{\cal Q}\rangle$, it is
convenient to write $\langle b'|\hat S|b\rangle$ as
$S^{BF}_{\Lambda',\Lambda}({\cal Q})\delta({\cal Q}-{\cal Q'})$, the
notation that has been already used in the above discussion. The
explicit form of the matrix elements $\langle s|b\rangle$ for the
unitary transformation is given in \cite{kokoouline04a,kokoouline04b}.

The $S^{BF}$ matrix is diagonal with respect to the rotational quantum
numbers $J_{\rm tot},\,K_{\rm tot}$ and $M_{\rm tot}$ of the whole
molecule and the continuous coordinate $\cal Q$. $J_{\rm tot},\,K_{\rm
tot}$ and $M_{\rm tot}$ are the angular momentum of the neutral molecule,
and the two projections of the angular momentum on the molecular symmetry
axis and on the space-fixed $z$-axis, respectively. To completely define
the BF basis functions $|b\rangle$, in addition to $\Lambda$ and $\cal Q$,
the rotational quantum numbers $J_{\rm tot},\,K_{\rm tot}$ and $M_{\rm
tot}$ must be also specified; for brevity they are omitted in the above
equation although their presence is implied.

To specify the vibrational states in the SF representation, we will use
the normal mode approximation, i.e. specifying ionic vibrational
eigenstates by the quantum numbers \{$v_1,v_2^{l_2}$\}. The rotational
part of the total wave function is specified by rotational quantum numbers
$J_{\rm tot}, J, K, M,l, m$, where $J, K$, and $M$ are the angular
momentum of the ion, and the two projections of the angular momentum on
the molecular symmetry axis and on the space-fixed $z$-axis, $m$ is the
projection of electronic angular momentum $l$ on the space-fixed 
$z$-axis. In the following, we will not specify any other conserved
quantum numbers that are the same in both bases, such as the total nuclear
spin and the irreducible representation of the total wavefunction.

The matrix $S^{SF}_{s';s}$ obtained by the above procedure does not yet
represent the physical scattering matrix \citep{aymar96,seaton66}. In
fact, it represents the actual scattering matrix $S^{phys}(E)$ only for
energies high enough such that {\it all} of the channels $|s\rangle$ are
open for electron escape, i.e. where the total energy of the system is
higher than the energy of the highest relevant ionization channel
$|s\rangle$. When at least one channel is closed, the physical scattering
matrix $S^{phys}(E)$ is obtained from $S^{SF}$ using the standard MQDT
channel-elimination formula (see Eq. (2.50) in \cite{aymar96} or Eq. (38)
in \cite{kokoouline04a}).

In terms of the energy-dependent scattering matrix $S^{phys}(E)$, the
cross-section for rovibrational (de-)excitation of the ion from the
initial state $|s\rangle$ is written (in atomic units, a.u.) as:
\begin{equation}
\label{eq:cs}
\sigma^{\rm RFT}_{s'\leftarrow s}(E_{el}) =
\frac{\pi}{2E_{el}(2J+1)}\sum_{J_{\rm tot}K_{\rm tot}}(2J_{\rm
tot}+1)|S_{s';s}^{(J_{\rm tot},
K_{\rm tot})}|^2,
\end{equation}
where $E_{el}$ is the relative kinetic energy of the ion and the electron
before the collision. In the above expression it has been assumed that the
initial $|s\rangle$ and final $|s'\rangle$ states are different.

\section{Rate constants for vibrational (de-)excitation}
\label{sec:vib_exc}

If one is not interested in the rotational structure of initial and final
vibrational states, the final cross-section (or thermal rate constant) has
to be averaged over the initial rotational levels and summed up over the
final rotational levels. This can be done using the full rovibrational
frame transformation technique described above. However, the cross-section
averaged over initial rotational levels and summed up over the final
rotational levels is very similar to the one obtained neglecting the
rotational structure of the ion. The cross-section obtained accounting for
the rotational structure has more resonances due to interaction between
rotational states but the averaged value is close to the value obtained
without the rotational structure. The quantity of interest in
astrophysical applications is the thermally-averaged rate constant.
Because the thermally-averaged rate is not sensitive to the position of
individual resonances in the energy-dependent cross-section, the
calculations with and without the rotational structure give the same
result.

\begin{figure}
\includegraphics[width=20pc]{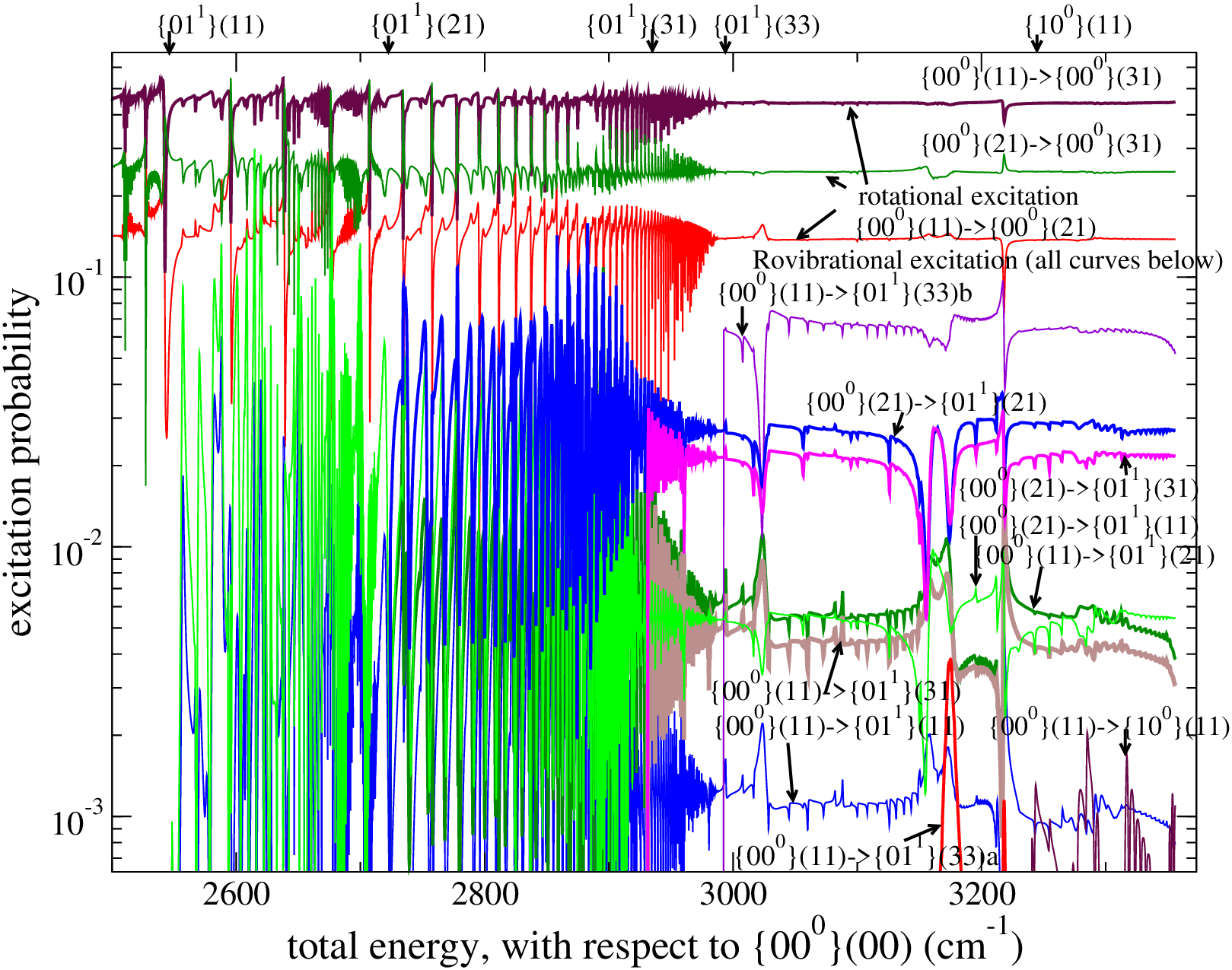}
\caption{(Color online) Probabilities of rovibrational excitation of the
H$_3^+$ ion calculated using the full rovibrational frame transformation.
Only transitions from the ground vibrational level $\{00^0\}$ are shown.
The $\{00^0\}\to \{01^1\}$ probabilities oscillate a lot below 3000
cm$^{-1}$ and become less energy-dependent above. The oscillations are due
to the strong rotational coupling between individual rotational levels of
the initial and final states of the ion. When averaged over the initial
and summed over the final rotational states and averaged over the
appropriate energy distribution, the resulting probabilities are similar
in magnitude to the probabilities shown in Fig.
\ref{fig:vibr_excitations}. The labels on top of the figure indicate
different rovibrational ionization limits. Notice that the zero
of energy in
the figure is set to the energy of the forbidden rovibrational level
$\{00^0\}(00)$.}
\label{fig:rovibr_excitations}
\end{figure}

\begin{figure}
\includegraphics[width=20pc]{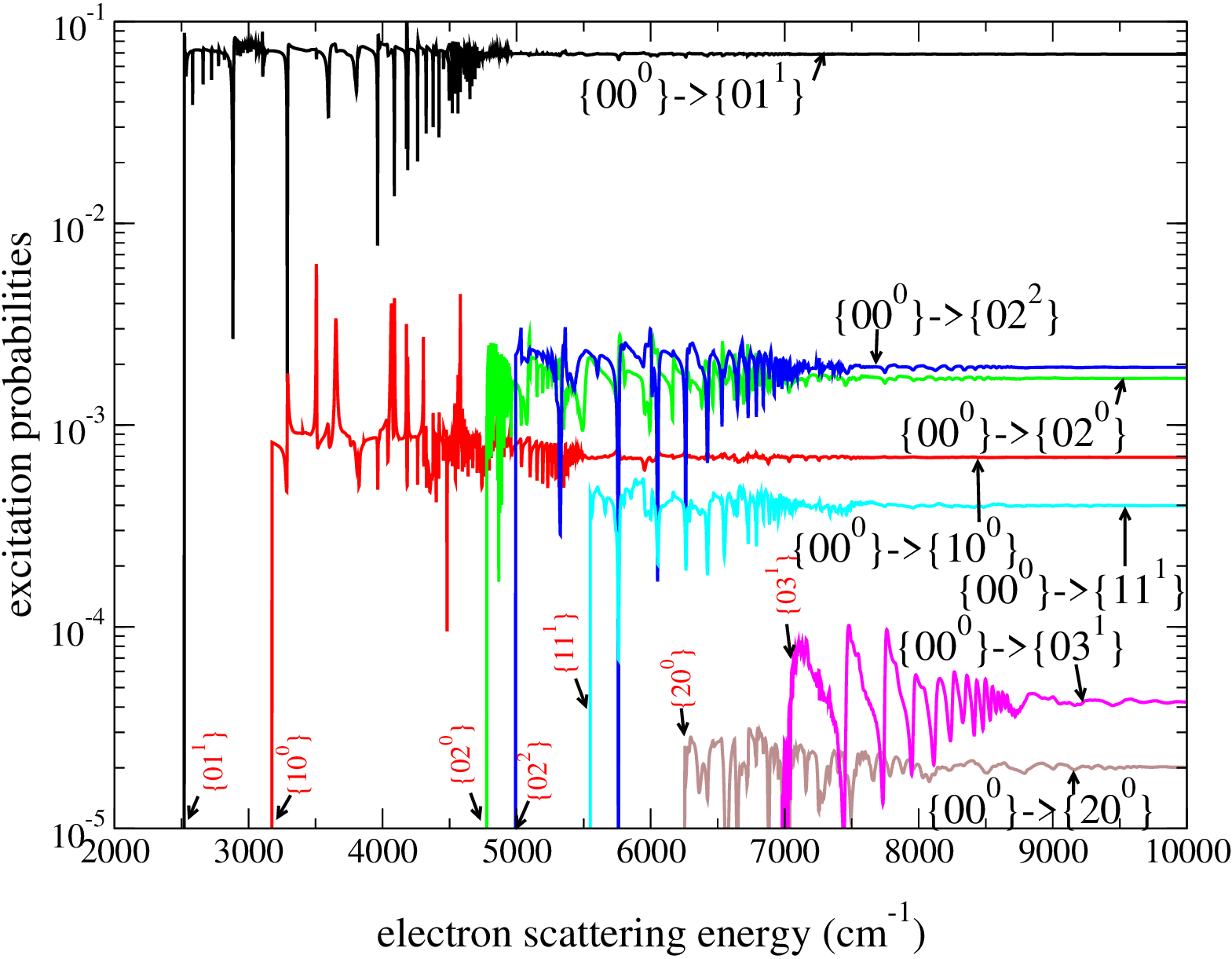}
\caption{(Color online) Probabilities of vibrational excitation from the
ground vibrational level $\{00^0\}$ to several excited vibrational levels
calculated using the vibrational frame transformation only. Energies of
vibrational thresholds are labeled with arrows and the corresponding
vibrational quantum numbers.}
\label{fig:vibr_excitations}
\end{figure}

Figs. \ref{fig:rovibr_excitations} and \ref{fig:vibr_excitations}
compare our calculations with and without rotational structure
included, for the probabilities of rovibrational and vibrational
(de-)excitation of the ion. Fig. \ref{fig:rovibr_excitations} shows in
detail the rovibrational transitions from the ground to the first
excited vibrational level $\{01^1\}$ with $J_{tot}=2$ in
para-H$_3^+$. In order to compare with the results on
Fig. \ref{fig:vibr_excitations}, one would need to take a sum over
final quanta and average over initial $J$ and $K$ and account for all
possible $J_{tot}$ and $K_{tot}$ similar as it is done in Eq.
\ref{eq:cs}. This would mean that to achieve a converged result at
reasonably high energy ($\sim$ 2000 cm$^{-1}$) calculations would be
needed for all $J_{tot}$ up to 10. This would require a tremendous
numerical effort if the fully quantum approach were to be applied.

\begin{figure}
\includegraphics[width=20pc]{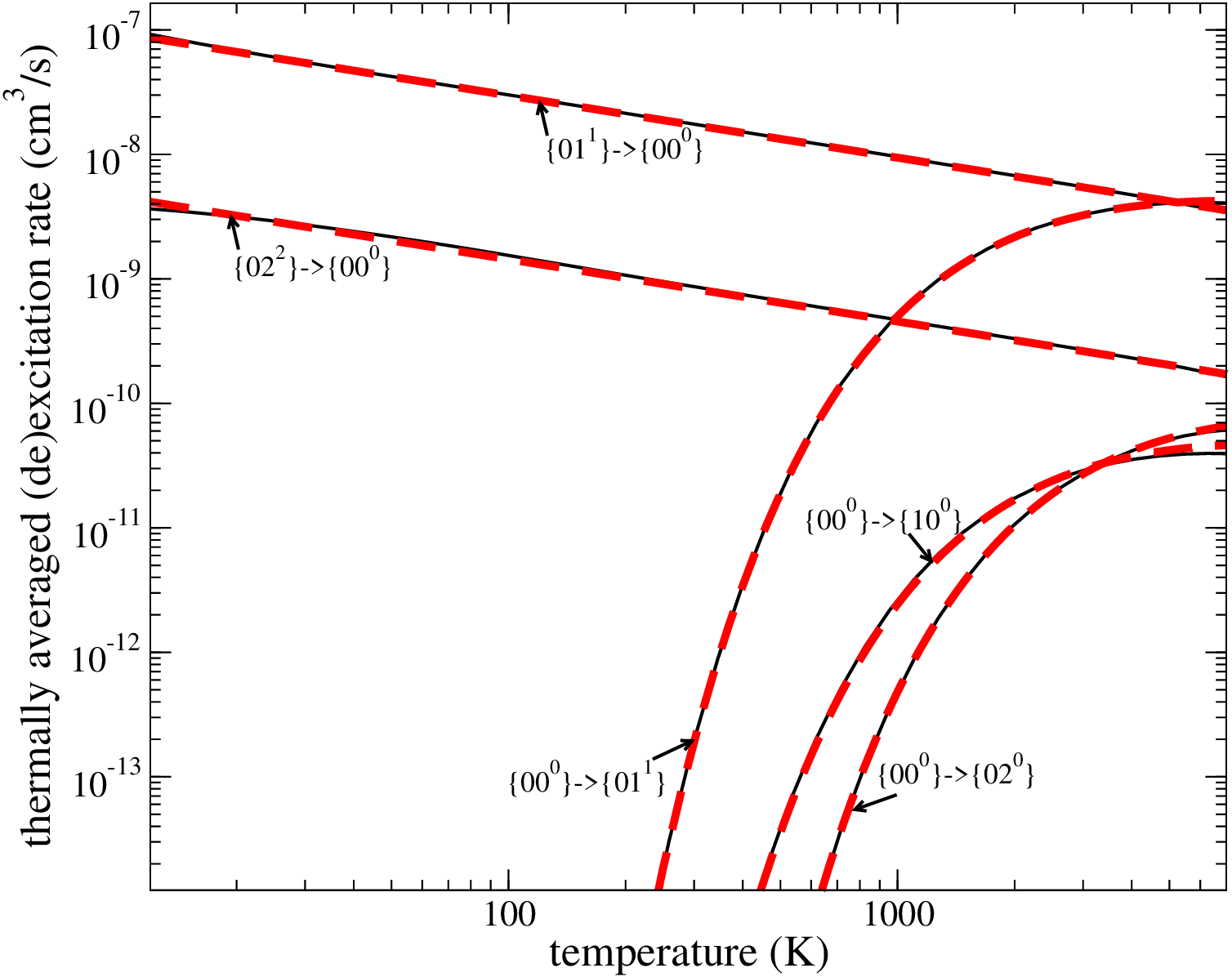}
\caption{(Color online) Thermally-averaged rate constants for several
(de-)excitation transitions obtained by direct integration using Eq.
(\ref{eq:ther_int}) (solid lines) and the approximate formula of Eq.
(\ref{eq:ther_appr}) (dashed line). The averaged probabilities for
vibrational (de-)excitations are listed in Table \ref{tab:1}.}
\label{fig:vibr_excitations_thermal_rates}
\end{figure}

Neglect of the rotational structure of the initial and final vibrational
state simplifies considerably the numerical calculation. The complete
rovibrational frame transformation of Eq. (\ref{eq:RVFT}) is reduced to
the vibrational frame transformation if the rotation is neglected, i.e it
is carried out using the following formula
\begin{equation}
\label{eq:Svv}
S^{SF}_{(v'\Lambda')(v\Lambda)}=\left\langle
v'|S^{BF}_{\Lambda';\Lambda}({\cal Q})|v\right\rangle ,
\end{equation}
where the brackets imply an integration over the vibrational
coordinates only. Many elements among
$S^{SF}_{(v'\Lambda')(v\Lambda)}$ are zero because of the symmetry of
vibrational wave functions and matrix elements
$S^{BF}_{\Lambda'\Lambda}({\cal Q})$. The vibrational (de-)excitation
$v\to v'$ cross-section obtained from the scattering matrix of Eq.
(\ref{eq:Svv}) should be averaged over $\Lambda$ and summed over
$\Lambda'$.

The thermally averaged rate constant $\alpha_{th}(T)$ (in a.u.) is
obtained from the energy-dependent cross-section $\sigma(E)$ as
\begin{eqnarray}
\label{eq:ther_int}
\alpha_{th}(T)=\frac{8\pi}{(2\pi kT)^{3/2}}\int_0^{\infty}\sigma(E_{el})e^{-\frac{E_{el}}{kT}}E_{el}dE_{el}\,,
\end{eqnarray}
where $T$ is the temperature. Temperature dependencies $\alpha_{th}(T)$ for
different (ro-)vibrational transitions $v\to v'$ obtained using Eq.
(\ref{eq:ther_int}) are shown in Fig.
\ref{fig:vibr_excitations_thermal_rates} as solid lines.

For further discussion, it is convenient to represent the cross-section
$\sigma(E_{el})$ in the form
\begin{eqnarray}
\sigma(E_{el})=\frac{\pi}{k^2}P(E_{el})\,,
\end{eqnarray}
where $k$ is the wave vector of the incident electron, $P(E_{el})$ is the
probability for vibrational (de-)excitation at collision energy $E_{el}$.
Figures \ref{fig:rovibr_excitations} and \ref{fig:vibr_excitations}
suggest that on average (here we mean a running average taken over a few
intervals between resonances), the probability behaves approximately as a
step function
\begin{eqnarray}
\label{eq:P_DR}
\langle P(E_{el})\rangle=P_0\ \theta(E_{el}-\Delta_{v',v})\,,
\end{eqnarray}
where $\Delta_{v'v}=E_{v'}-E_{v}$ is the threshold energy for
(ro-)vibrational excitation (if $E_{v'}-E_{v}>0$); $\Delta_{v'v}=0$ for
de-excitation (if $E_{v'}-E_{v}<0$), $\theta$ is the Heaviside function,
$P_0$ is a constant. The above approximation for the probability is
accurate enough to calculate the thermally-averaged rate constant that is
not sensitive to the detailed resonance structure of the energy-dependent
(de-)excitation cross-section. Using Eq. (\ref{eq:P_DR}) the thermally
rate constant of Eq. (\ref{eq:ther_int}) becomes
\begin{eqnarray}
\label{eq:ther_appr}
\alpha_{th}(T)=\sqrt{\frac{2\pi}{kT}}e^{-\frac{\Delta_{v'v}}{kT}}P_0\,.
\end{eqnarray}
The above formula with only one parameter $P_0$ provides a very good
approximation for the actual thermally averaged rate constant. It is
demonstrated in Fig.  \ref{fig:vibr_excitations_thermal_rates} that
compares the thermal rate constants for different (de-)excitation
transitions obtained with the direct numerical integration using Eq.
(\ref{eq:ther_int}) and with the approximate formula of Eq.
(\ref{eq:ther_appr}). Therefore, for practical applications, it is
convenient to provide just averaged probabilities $P_0$ and the
energies of vibrational thresholds $E_v$ for each pair of vibrational
(de-)excitations. These parameters are listed in Table~\ref{tab:1} for
all combinations of the first eight vibrational states of
H$_3^+$. {\bf Note that the conversion factor from a.u. to cm$^3$/s is
  6.126$\times 10^{-9}$.}

\begin{table*}
 \centering
 \begin{minipage}{155mm}
\begin{tabular}{|l|l|l|l|l|l|r|r|r|}
\hline
          &      $\{00^0\}$0  &   $\{01^1\}$2521   &   $\{10^0\}$3178 &   $\{02^0\}$4778  &   $\{02^2\}$4998  &   $\{11^1\}$5554  &   $\{20^0\}$6262&    $\{03^1\}$7006\\ \hline\hline
$\{00^0\}$&                   & 3.7$\times10^{-2}$& 8.5$\times10^{-4}$& 1.5$\times10^{-3}$& 1.1$\times10^{-3}$& 2.0$\times10^{-4}$& 2.3$\times10^{-5}$& 2.1$\times10^{-5}$\\
$\{01^1\}$& 6.9$\times10^{-2}$&                   & 1.2$\times10^{-2}$& 6.0$\times10^{-2}$& 6.4$\times10^{-2}$& 3.3$\times10^{-3}$& 2.2$\times10^{-4}$& 3.6$\times10^{-3}$\\
$\{10^0\}$& 8.0$\times10^{-4}$& 5.7$\times10^{-3}$&                   & 2.0$\times10^{-3}$& 5.9$\times10^{-4}$& 3.5$\times10^{-2}$& 1.4$\times10^{-3}$& 2.4$\times10^{-4}$\\
$\{02^0\}$& 1.7$\times10^{-3}$& 2.9$\times10^{-2}$& 1.9$\times10^{-3}$&                   & 1.2$\times10^{-2}$& 3.5$\times10^{-3}$& 8.8$\times10^{-5}$& 4.3$\times10^{-2}$\\
$\{02^2\}$& 1.9$\times10^{-3}$& 6.3$\times10^{-2}$& 1.4$\times10^{-3}$& 3.0$\times10^{-2}$&                   & 1.4$\times10^{-2}$& 1.7$\times10^{-4}$& 2.6$\times10^{-2}$\\
$\{11^1\}$& 3.9$\times10^{-4}$& 2.8$\times10^{-3}$& 7.1$\times10^{-2}$& 6.3$\times10^{-3}$& 1.3$\times10^{-2}$&                   & 1.8$\times10^{-2}$& 2.3$\times10^{-3}$\\
$\{20^0\}$& 2.0$\times10^{-5}$& 1.0$\times10^{-4}$& 1.3$\times10^{-3}$& 6.7$\times10^{-5}$& 1.0$\times10^{-4}$& 1.1$\times10^{-2}$&                   & 1.2$\times10^{-5}$\\
$\{03^1\}$& 4.3$\times10^{-5}$& 4.1$\times10^{-3}$& 6.3$\times10^{-4}$& 1.0$\times10^{-1}$& 2.8$\times10^{-2}$& 2.2$\times10^{-3}$& 2.8$\times10^{-5}$& \\
\hline
\end{tabular}
\vspace{0.3cm}
\caption{Parameters $P_0$ for several vibrational transitions that can be used in the approximate formula, Eq. (\ref{eq:ther_appr}), for thermally-averaged rate constants. Initial states, $v$ are given in the upper row, final states $v'$ -- in the left column. The upper row specifies also the energies $E_v$ (in cm$^{-1}$) of vibrational levels. The probabilities $P_0$ are obtained by fitting numerical dependencies obtained by a direct integration of Eq. (\ref{eq:ther_int}). That is why $P_0(v'\to v)$ is not exactly equal to $P_0(v\to v')$. Notice that multiplicity factors of doubly degenerate vibrational states $E$ are taken into account in the probabilities. For example, $P_0(v\to v')\approx 2P_0(v'\to v)$, if the vibrational states $v$ and $v'$ are the states of the $A_1$ and $E$ irreducible representations correspondingly.}
\label{tab:1}
\end{minipage}
\end{table*}

\section{Rate constants for rotational (de-)excitation}
\label{sec:rot_exc}

\begin{figure}
\includegraphics[width=20pc]{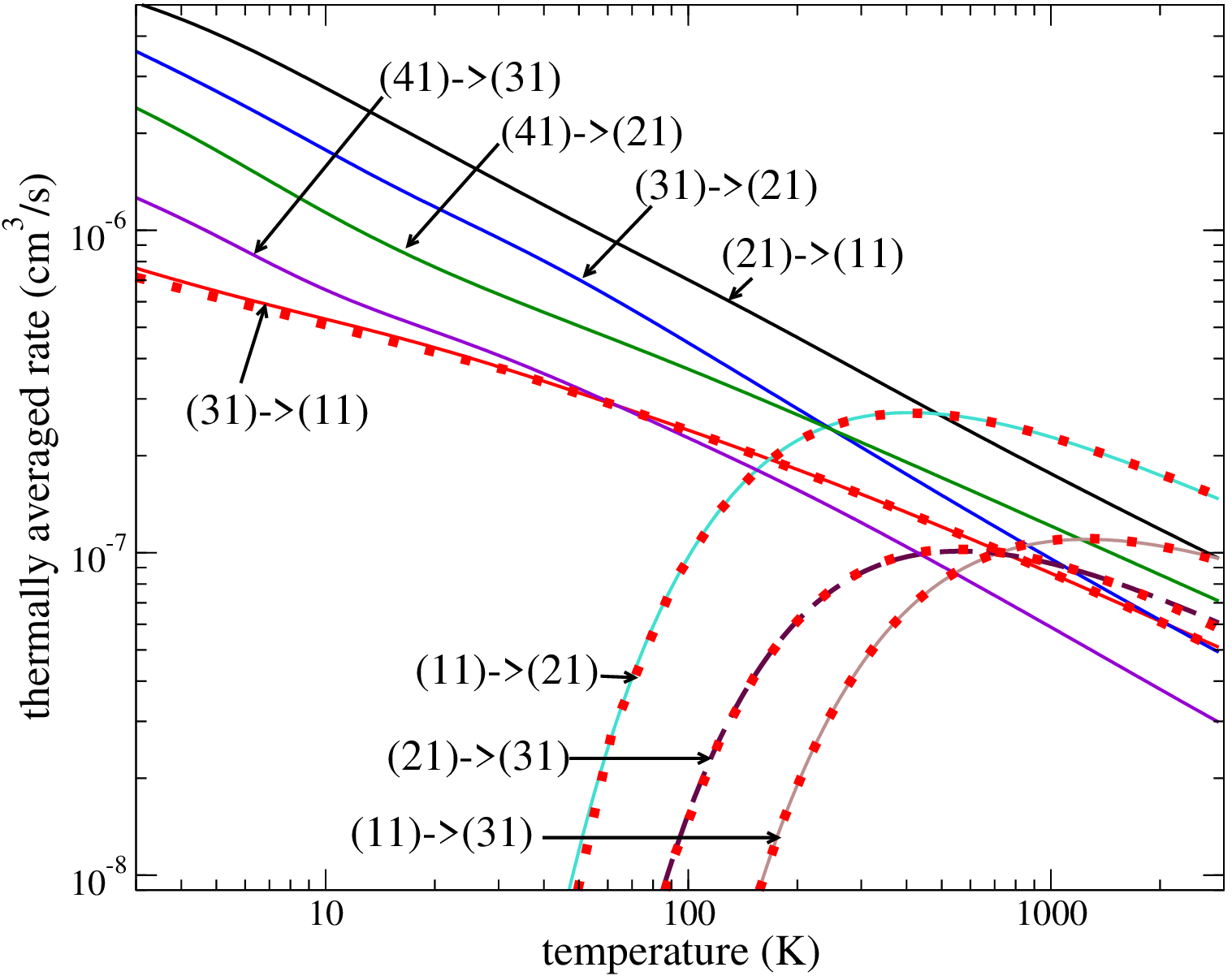}
\caption{(Color online) Thermally-averaged rate constants for several
rotational (de-)excitation transitions $(JK)\to(J'K')$ of the H$_3^+$ ion
(solid lines). The vibrational level, $\{00^0\}$, is the same in the
initial and final state of the ion. The dotted lines show a few examples
of the numerical fit using Eq. (\ref{eq:rot_exc_fit}).}
\label{fig:rot_exc}
\end{figure}

If the temperature $T$ of the H$_3^++e^-$ plasma is not very high,
such that only the ground vibrational level $\{00^0\}$ of H$_3^+$ is
significantly populated, knowledge of rate constants for transitions
$r\to r'$ between individual rotational levels $r$ and $r'$ of
$\{00^0\}$ may be important for the analysis of experimental or
astronomical spectra. For this purpose, we have made a detailed
analysis of transitions between individual rotational states of the
ground vibrational level. The calculation of rotational
(de-)excitation rate constants was carried out using the cross-section
of Eq. (\ref{eq:cs}) and numerical integration of
Eq. (\ref{eq:ther_int}).  Examples of the thermally-averaged rate
constants for the rotational (de-)excitation are shown in
Fig. \ref{fig:rot_exc}.

 \begin{figure}
\includegraphics[width=20pc]{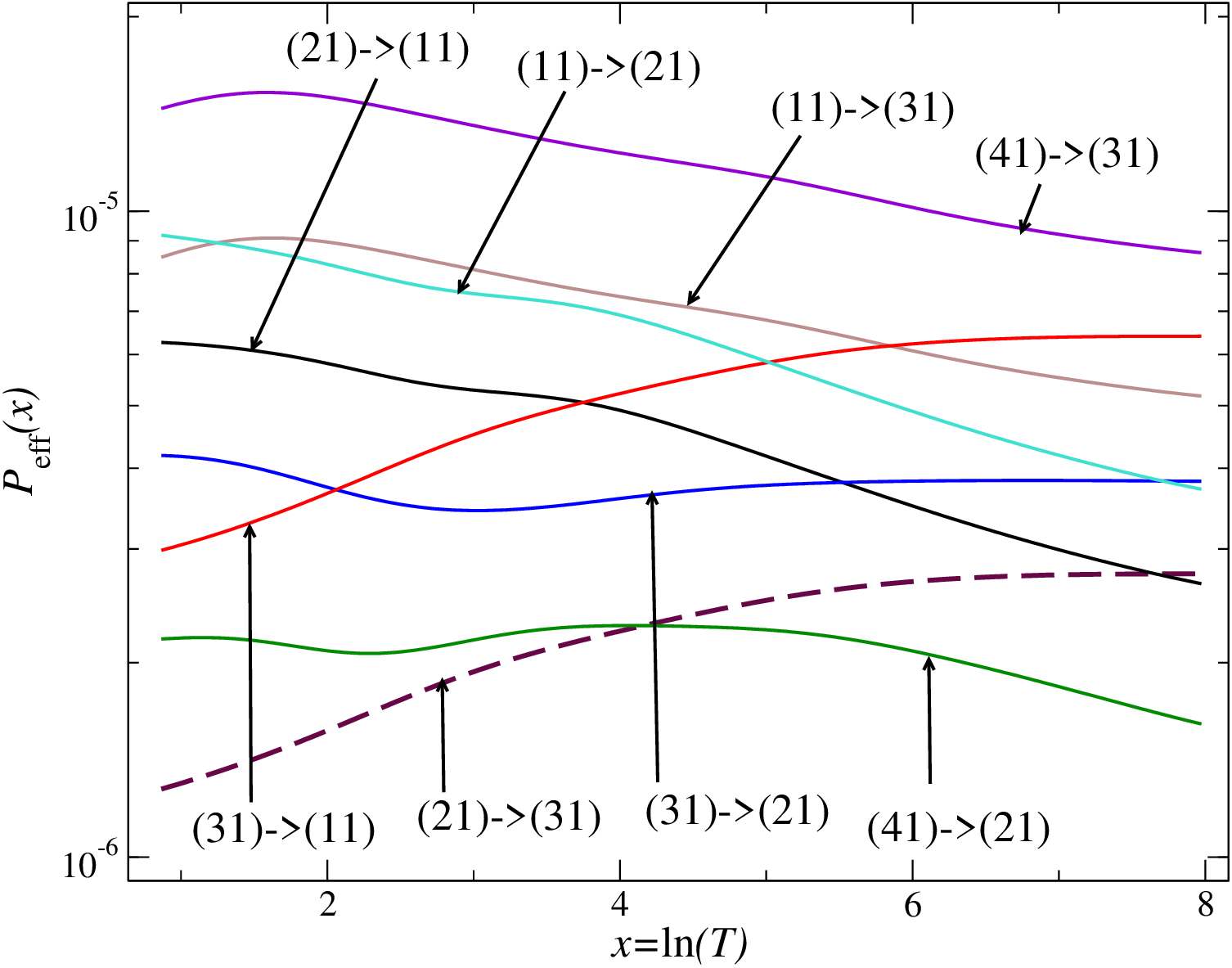}
\caption{(Color online) $P_{\rm eff}
=\sqrt{T}\exp\left({{\Delta_{r',r}}/{T}}\right)\alpha\left[J'K'\leftarrow
JK\right]$. Note that  $P_{\rm eff}$ is only weakly dependent on
temperature. It is used to obtain a cubic polynomial fit $P_m(\ln(T))$ in
Eq. (\ref{eq:rot_exc_fit}) for each transition $(JK)\to (J'K')$ with four
parameters that are listed in Tables  \ref{tab:2}, \ref{tab:3},
\ref{tab:4}, and \ref{tab:5}. }
\label{fig:rot_exc_reduced}
\end{figure}

\begin{table*}
 \centering
 \begin{minipage}{125mm}
\begin{tabular}{|r|r|r|r|r|r|r|r|}
\hline
	& (11)-(21) 	&(11)-(31) 	&  (21)-(31) 	&(21)-(41)	&(31)-(41)&    (31)-(51) &   (41)-(51)\\ \hline\hline
$\Delta_{r'r}$(K)&249         & 619  & 370 & 858& 488& 1087 & 600 \\
\hline
$a_0$	& 1.51e-5	& 2.29e-6	&9.47e-6	&1.01e-5	&3.17e-6	&7.04e-6	&3.24e-6	\\
	&8.81e-6	&9.74e-7	&6.33e-6	&5.18e-6	&2.20e-6	&4.24e-6	&2.41e-6	\\ 
\hline
$a_1$	& 4.6e-7	& 6.13e-7	&-3.76e-7     	&-2.64e-6	&-3.43e-7	&7.62e-8	&-4.56e-7	\\
	&4.08e-7	&2.72e-7	&-2.44e-8	&-1.21e-6	&-1.17e-7	&1.87e-7	&-2.31e-7	\\ 
\hline
$a_2$	&-3.96e-7 	&7.16e-8	&-1.14e-7      	&5.77e-7	&1.10e-7	&-1.82e-8	&4.54e-8	\\
	&-2.62e-7	&2.83e-8	&-1.25e-7	&2.74e-7	&5.81e-8	&-3.69e-8	&1.1e-8	\\ 
\hline
$a_3$	& 3.00e-8	&-1.07e-8	&8.57e-9       	&3.78e-8	&-1.09e-8	&7.6e-10	&-1.94e-9	\\
	&1.94e-8	&-4.42e-9	&8.73e-9	&-1.83e-8	&6.85e-9	&1.99e-9	&-2.3e-11	\\ 
\hline
\end{tabular}
\vspace{0.3cm}
\caption{Parameters $a_0,\ a_1,\ a_2,\ a_3$ of the fit polynomial $P_m(x)$ of Eq. (\ref{eq:rot_exc_fit}) for several transitions between rotational states of the $E''$ irreducible representation of the coordinate part of the ion-electron system. The vibrational level of the ion is the same in the initial and final states. The upper line specifies the pairs $(J_1K_1)-(J_2K_2)$ of rotational states for which the parameters are fit.  For convenience, we also specify (second line of the table) the threshold energy $\Delta_{r'r}$
for each transition. All but the first rows in the table have two numbers:  The upper number in each cell corresponds to the transition $(J_1K_1)\to(J_2K_2)$, the lower number corresponds to the reverse transition $(J_1K_1)\leftarrow (J_2K_2)$.}
\label{tab:2}
\end{minipage}
\end{table*}

\begin{table*}
 \centering
 \begin{minipage}{110mm}
\begin{tabular}{|r|r|r|r|r|r|r|}
\hline
&(22)-(32) 	&(22)-(42) &  (32)-(42) 	&(32)-(52)	&(42)-(52)&    (44)-(54) \\ \hline\hline
$\Delta_{r'r}$(K)&   372    & 862 & 490 &1092 & 602 & 614\\
\hline
$a_0$	& 1.39e-5	&3.69e-6	&8.36e-6	&5.94e-6	&9.79e-6	&1.07e5	\\
	&9.38e-6	&1.87e-6	&5.99e-6	&3.48e-6	&7.42e-6	&8.01e-6	\\ 
\hline
$a_1$	&-2.73e-6	&5.97e-8	& -1.34e-6   	&-5.51e-7	&-2.51e-6	&-7.78e-8	\\
	&-1.66e-6	&1.37e-7	&-7.52e-7	&-1.85e-7	&-1.73e-6	&3.53e-7	\\ 
\hline
$a_2$	& 5.98e-7	&-2.38e-8	&4.05e-7      	&8.42e-8	&4.69e-7	&-1.77e-7	\\
	&3.73e-7	&-3.22e-8	&2.63e-7	&2.32e-8	&3.25e-7	&-2.22e-7	\\ 
\hline
$a_3$	&-4.35e-8	&1.53e-9	& -3.55e-8      &-4.e-9		&-3.06e-8	&1.6e-8	\\
	&-2.79e-8	&1.98e-9	&-2.46e-8	&-7.4e-10	&-2.16e-8	&1.77e-8	\\ 
\hline
\end{tabular}
\vspace{0.3cm}
\caption{Parameters $a_i$  for several transitions between rotational states of the $E'$ irreducible representation of the ionic wave function. See Table \ref{tab:2} for details.}
\label{tab:3}
\end{minipage}
\end{table*}

As is evident from the figure, the rotational rate constants behave
approximately according to Eq. (\ref{eq:ther_appr}), where $\Delta_{v'v}$
should be replaced with the rotational threshold energy, 
$\Delta_{r'r}=(E_{r'}-E_{r})\theta(E_{r'}-E_{r})$. However, there is a
weak departure from the dependence of Eq.  (\ref{eq:ther_appr}). It is
clearer in Fig. \ref{fig:rot_exc_reduced}, where we plotted an
``effective'' value $P_{\rm eff}=\alpha_{r'\leftarrow
r}(T)\sqrt{T}\exp\left({\frac{\Delta_{r'r}}{T}}\right)$ of the parameter
$P_0$ as a function of $\ln(T)$. Notice that on average, the quantities
$P_{\rm eff}\left[(J'K')\leftarrow (JK)\right]$ and $P_{\rm
eff}[(JK)\leftarrow (J'K')]$ for the two opposite processes are related by
the  principle of detailed balance
\begin{equation}
 (2J+1)P_{\rm eff}\left[(J'K')\leftarrow (JK)\right]\approx (2J'+1)P_{\rm
eff}\left[(JK)\leftarrow (J'K')\right]\,,
\end{equation}
because $P_{\rm eff}(x)$ represents the thermally averaged probability of
the rotational transition per one electron-ion collision (see also Eq.
(\ref{eq:cs})).  In order to simplify eventual applications of the
calculated numerical constants $\alpha_{r',r}(T)$, we fitted the 
numerical rate constants representing the ``effective'' value of $P_0$ in
Eq. (\ref{eq:ther_appr}) to a cubic polynomial of $\ln(T)$. For this we used
the following analytical interpolation formula for $\alpha_{r',r}(T)$
\begin{eqnarray}
\label{eq:rot_exc_fit}
\alpha_{r'\leftarrow
r}(T)=\frac{1}{\sqrt{T}}\exp\left({-\frac{\Delta_{r'r}}{T}}\right) P_m(x)\
\textrm{, where}\,\nonumber\\
P_m(x)=a_0+a_1x+a_2x^2+a_3x^3\ \textrm{and}\\
x=\ln(T)\,,\nonumber
\end{eqnarray}
where the constants $a_i\ (i=0,1,2,3)$ are obtained for each individual
transition $r\to r'$ from a numerical fit. The quantity $P_m(x)$ in the above equation has a meaning of (de--~)excitation probability that varies weakly with energy. We used the subscript $m$ to distinguish it from $P_{\rm eff}$ (that is energy-independent) and to stress that $P_m(x)$ is given by a model polynomial. The  constants obtained, $a_i$,
are listed in Tables \ref{tab:2}, \ref{tab:3}, \ref{tab:4}, and
\ref{tab:5} for several combinations of initial $r=(J,K)$ and final
$r'=(J'K')$ rotational states of H$_3^+$.   The numerical values of $a_i$
listed in the table are such that, when plugged into Eq.
(\ref{eq:rot_exc_fit}), they give rate constants in units of cm$^3/$s.
Temperatures in the calculation of $x=\ln(T)$ should be in K.  Notice that
$(2J_1+1)a_0[1\to2]\approx(2J_2+1)a_0[1\leftarrow2] $ due to the  principle
of detailed
balance.

\begin{table}
\begin{center}
\begin{tabular}{|r|r|r|}
\hline
&      (10)-(30) 	&(30)-(50) \\ \hline\hline
$\Delta_{r'r}$(K)&  619              & 1085   \\
\hline
$a_0$	& 9.54e-6	&6.01e-6		\\
	&3.85e-6	&3.58e-6		\\ 
\hline
$a_1$	&-3.87e-7	&4.14e-8		\\
	&-3.32e-8	&1.64e-7		\\ 
\hline
$a_2$	& 1.43e-7	&1.04e-7	\\
	&3.71e-8	&4.12e-8		\\ 
\hline
$a_3$	&-1.155e-8	&-1.06e-8	\\
	&-3.50e-9	&-5.23e-9		\\ 
\hline
\end{tabular}
\vspace{0.3cm}
\caption{The table gives parameters $a_i$  for several transitions between rotational states of the $A_2'$ irreducible representation of the ionic wave function. See Table \ref{tab:2} for details.}
\label{tab:4}
\end{center}
\end{table}

\begin{table}
\begin{center}
\begin{tabular}{|r|r|r|r|}
\hline
&      (33)-(43) 	&(33)-(53) 	&  (43)-(53) \\ \hline\hline
$\Delta_{r'r}$(K)&  494              &  1101 &  607 \\
\hline
$a_0$	& 1.39e-5	&2.16e-6	&1.26e-5	\\
	&9.92e-6	&1.31e-6	&9.84e-6	\\ 
\hline
$a_1$	&-1.34e-6	&7.37e-8	&-2.07e-6 	\\
	&-5.56e-7	&8.18e-8	&-1.41e-6	\\ 
\hline
$a_2$	& 4.56e-8	&-1.81e-8	&3.86e-7    	\\
	&-5.41e-8	&-1.79e-8	&2.65e-7	\\ 
\hline
$a_3$	&3.38e-9	&1.09e-9	& -2.75e-8     	\\
	&7.95e-9	&1.07e-9	&-1.95e-8	\\ 
\hline
\end{tabular}
\vspace{0.3cm}
\caption{The table gives parameters $a_i$  for several transitions between rotational states of the $A_2''$ irreducible representation of the ionic wave function. See Table \ref{tab:2} for details.}
\label{tab:5}
\end{center}
\end{table}

\section{Astrophysical implications}
\label{sec:astro}

\begin{table*}
 \centering
 \begin{minipage}{100mm}
\begin{tabular}{|c|c|c|c|c|c|c|c|}
\hline 
$T$ & $\{01^1\}$ & $\{10^0\}$ & $\{02^0\}$ & $\{02^2\}$ &
$\{11^1\}$ & $\{20^0\}$ & $\{03^1\}$ \\
 \hline
\hline 
10 & 1.3(9) & 2.4(7) & 3.1(9) & 2.6(9) & 9.5(8) & 4.8(7) & 1.3(9) \\ 
100 & 4.0(9) & 7.7(7) & 9.7(9) & 8.1(9) & 3.0(9) & 1.5(8) & 4.2(9) \\ 
1000 & 1.1(10) & 2.0(8) & 2.3(10) & 2.1(10) & 8.9(9) & 4.8(8) & 1.3(10)\\ 
\hline
\end{tabular}
\caption{Critical electron density, $n_{\rm cr}$ in cm$^{-3}$, at 10,
  100 and 1000~K, for vibrational levels of H$_3^+$. Powers of 10 are
  given in parentheses.}
\label{tab:6}
\end{minipage}
\end{table*}

The vibrational and rotational distribution of H$_3^+$ ions in
interstellar space, planetary ionospheres and in laboratory is
determined by the competition between radiative and collisional
processes. While the ionization level is generally low in
astrophysical plasmas, electrons can still play a role in the
molecular excitation because electron-impact rates exceed those for
excitation by neutrals (H, He, H$_2$) by several orders of
magnitude. Detailed excitation models, including collisional data for
all relevant colliders, are therefore required to derive reliable
column densities from astronomical observations.

An important concept in this context is the so-called critical
density, $n_{cr}$, which is defined as the density at which the
collisional rate is equal to the spontaneous radiative rate. The usual
definition refers to a specific transition in a two-level
approach. For a multi-level system, neglecting opacity effects, a
practical definition is to refer to a specific level $s$:
\begin{equation}
\label{eq:ncr}
n_{cr}(s, T)=\frac{\sum_{s'}A(s\rightarrow s')}
{\sum_{s'}\alpha(s\rightarrow s', T)},
\end{equation}
where $A(s\rightarrow s')$ are the Einstein coefficients for
spontaneous emission, $\alpha(s\rightarrow s', T)$ are the collisional
rates and the sums run over all possible transitions $s\rightarrow
s'$. Considering electron collisions only, these latter will maintain
the level $s$ in local thermodynamic equilibrium (LTE) for electron
densities $n_e\gg n_{cr}(s, T)$, while deviations from LTE including
population inversions are expected for densities $n_e\lesssim
n_{cr}(s, T)$. For $n_e\ll n_{cr}(s, T)$, electron collisions will be
negligible. Non-LTE effects caused by H$_2$ collisions have been
investigated both in interstellar clouds \citep{oka04,oka05} and in
the Jovian atmosphere \citep{melin05}.  We note in this context
  that microcanonical statistical calculations have been performed
  recently to estimate thermal state-to-state rate coefficients for
  the H$_3^+$+H$_2$ reaction and its deuterated variants
  \citep{park07,hugo09}. Electron-impact rotational excitation has
been considered by \cite{faure06b} but, to the best of our knowledge,
electron-impact vibrational excitation has been so far ignored in
non-LTE modelling.

Eq.~(\ref{eq:ncr}) was computed with the vibrational and rotational
rates presented in the previous sections. Einstein $A$ coefficients
were taken from \cite{dinelli92a,dinelli92b} for vibrational
transitions and from \cite{pan86} for rotational transitions. Results
are presented in Tables~\ref{tab:6} and \ref{tab:7}. It can be noticed
that critical densities for vibrational and rotational levels differ
by typically 6 orders of magnitude: they range between 10$^7$ and
10$^{10}$cm$^{-3}$ for the former and between 10$^{-1}$ and
10$^{4}$cm$^{-3}$ for the latter. In the diffuse interstellar medium,
the electron density is $\sim$0.1~cm$^{-3}$ \citep{black91} while it
can reach about 10$^6$~cm$^{-3}$ in planetary atmospheres (see for
example \citealt{lystrup08}). As a result, non-LTE rotational
populations are expected in interstellar clouds whereas rotational
levels should be at or close to LTE in the jovian atmosphere. A
non-thermal rotational distribution of H$_3^+$ was actually observed
towards galactic center clouds where the metastable (3, 3) level has a
population comparable to that in (1, 1) despite being 361.5~K higher
\citep{oka05}. On the other hand, electrons are expected to be
negligible in vibrationally exciting H$_3^+$ in the interstellar
medium but they could establish a non-LTE vibrational population of
H$_3^+$ in planetary environments, as observed in the jovian
thermosphere \citep{kim92}.

Table~\ref{tab:6} shows that the two levels $\{10^0\}$ and
$\{20^0\}$ have critical densities significantly lower than the other
levels. This directly reflects the lower Einstein $A$ coefficients of
$\sim$1~s$^{-1}$ \citep{dinelli92a}. We note that the values are
however not low enough to ensure LTE vibrational population in
planetary atmospheres where $n_e<10^7$~cm$^{-3}$.

\begin{table*}
 \centering
 \begin{minipage}{185mm}
\begin{tabular}{|c|c|c|c|c||c|c||c|c||c|c|c|c|c|c|}
\hline
$T$  & (21)    & (31)   &  (41)  &   (51)  & (30)    &  (50)&   (43)   &  (53)& (22)    &  (32)  &   (42)  &  (52) &   (44)  &   (54) \\ \hline\hline
10     & 2.5(-1) & 2.7(1) & 4.6(2) &  2.4(3)& 3.6(1) & 2.2(3)& 3.6(1) & 2.2(3)& 3.9(15) & 1.3(1) & 1.2(2)  & 1.1(3)  &  6.4(23) &1.3(1)\\
100    & 9.6(-1) & 9.4(1) & 1.2(3) & 7.9(3)& 1.1(2)  & 6.0(3) & 6.7(1) & 2.0(1) & 3.6(1)  & 4.3(1) & 3.8(2)  & 3.8(3)  &  2.3(0)  &4.8(1) \\
1000   & 1.9(0)  & 1.5(2) & 2.5(3) & 2.7(4)& 2.1(2)  & 1.8(4) & 1.4(2) & 7.0(1)& 3.5(0)  & 7.7(1) & 9.4(2)  & 1.3(4)  &  3.5(-2) &1.8(2)\\
\hline
\end{tabular}
\caption{Critical electron density, $n_{\rm cr}$ in cm$^{-3}$, at 10,
  100 and 1000~K, for several rotational levels of  H$_3^+$ (the levels are  grouped according to the corresponding irreducible representations). Powers of 10 are given in parentheses.}
\label{tab:7}
\end{minipage}
\end{table*}

Table~\ref{tab:7} shows that at 10~K the levels (22) and (44) have
much higher critical densities than the other levels. This actually
reflects the fact that these two levels can depopulate collisionally
through {\it excitation} only since rates for rotational transitions
with $\Delta K \ne 0$ are null within our treatment. These rates, as
those with $|\Delta J|>2$, were actually estimated by
\cite{faure03} and \cite{faure06a} and were found to be 3 to 4 orders of
magnitude smaller than those with $\Delta J=\pm1, \pm 2$ and $\Delta
K=0$. The (44) level is particularly interesting because an
astrophysical maser is predicted in the (44)$\rightarrow$(31)
transition of H$_3^+$ \citep{black00}.  It should be noted that the
selection rules for the (forbidden) rotational radiative transitions
are $\Delta J=0, \pm 1$ and $\Delta K=\pm 3$ \citep{pan86,miller88}. Critical
densities suggest that electrons might contribute, in some
environments, to create and maintain the necessary population
inversion. Note that in Table~\ref{tab:7} the (33) level is not listed
since it is metastable. In this case, the concept of critical density
is meaningless. Finally we emphasize that critical densities provide
guidance at the order of magnitude level and that a detailed non-LTE
modelling, including all relevant colliders, is necessary to properly
quantify deviations from LTE.

\section{Conclusion}
\label{sec:concl}

In this study we have performed calculations of thermally averaged
rate constants for rotational and vibrational transitions in H$_3^+$
caused by an electron impact. The calculations were made from the
first principles using the quantum defect approach. The rotational
rate constants are calculated for the ground vibrational level of the
ion in the initial and final states. The rate constants for
transitions between different vibrational levels are calculated
neglecting the rotational substructure of each vibrational level,
which corresponds to averaging over initial rotational states and
summing over the final rotational states. The obtained
thermally-averaged rate constants are well described by the analytical
formula of Eq. (\ref{eq:ther_appr}) with the parameter $P_0$, that can
be considered as temperature independent for vibrational transitions,
and weakly dependent on temperature for the rotational
transitions. For the rotational transitions, we have made a numerical
fit of the parameter by a cubic polynomial. The numerical values of
the fitting procedure are provided in Tables~1-5. The presented
thermally-averaged rate constants can be useful in interpretation of
hydrogen-dominated plasma experiments as well as for modelling
interstellar clouds and planetary atmospheres, where the H$_3^+$ ion
is present. The computation of critical densities suggests in
particular that electrons could establish non-LTE rotational
populations of H$_3^+$ in diffuse interstellar clouds and non-LTE
vibrational populations in planetary atmospheres.

This work has been supported by the National Science Foundation under Grant No.  PHY-0855622, by the Department of Energy, 
Office of Science, and by an allocation of NERSC supercomputing resources. Support by the French CNRS National Program "Physique et Chimie du Milieu Interstellaire" and the  {\it R\'eseau th\'ematique de recherches avanc\'ees "Triangle de la Physique"} is acknowledged.
\bibliographystyle{apj}

\bsp

\label{lastpage}

\end{document}